\documentclass[aps,prb,showpacs,preprint,amsmath,amssymb]{revtex4}
\usepackage{graphicx}
\usepackage{dcolumn}
\usepackage{bm}
\usepackage{float}

\begin{document}

\title{Phonon Knudsen flow in nanostructured semiconductor systems}
\author{E. Ziambaras and P. Hyldgaard}
\affiliation{Department of Applied Physics, Chalmers University of
Technology, SE--412 96 G\"oteborg, Sweden}
\date{\today}

\begin{abstract}
We determine the size effect on the lattice thermal conductivity of
nanoscale wire and multilayer structures formed in and by some typical
semiconductor materials, using the Boltzmann transport equation and focusing 
on the Knudsen flow effect.  For both types of nanostructured systems we find 
that the phonon transport is reduced significantly below the bulk value by
boundary scattering off interface defects and/or interface modes. 
The Knudsen flow effects are important for almost all types of 
semiconductor nanostructures but we find them most pronounced in Si 
and SiC systems due to the very large phonon mean-free paths.  
We apply and test our
wire thermal-transport results to recent measurements on Si
nanowires. We further investigate and predict size effects in 
typical multilayered SiC nanostructures, for example, a 
doped-SiC/SiC/SiO$_2$ layered structure that could define the 
transport channel in a nanosize transistor.  
Here the phonon-interface scattering produces a heterostructure 
thermal conductivity smaller than what is predicted in a traditional 
heat-transport calculation, suggesting a breakdown of the traditional 
Fourier analysis even at room temperatures.  Finally, we show that the effective
thermal transport in a SiC/SiO$_2$ heterostructure is sensitive to the oxide
depth and could thus be used as an in-situ probe of the SiC oxidation progress.
\end{abstract}

\maketitle

\section{Introduction}
Thermal transport in nanostructured semiconductor systems is
important for the continued development of nanoscale 
electronics.\cite{mahan2,cahill} A number of recent 
experimental\cite{deyu,yao,tighe,lee,Schwab,chen1,cahill2} and 
theoretical\cite{chen00,per1,ren,rennert,baladin,wire,wiretheory,X,richardson,zou,volz2,mingowire,luxiang,cher,X2,X3,barrat,chenperp,per2,per3,broidoperp,tamura,chen2,volz1,bies,chen0}
studies contributes to an emerging understanding of interesting science
issues and potential technology problems.  Opto- and power-electronics 
devices generate significant heat that must be dissipated to avoid 
degradation of the device performance,\cite{cahill,cl_tien} and similar
problems increasingly affect also more standard electronics applications.

The thermal transport in semiconductor structures is dominated by phonons 
and measurements on bulk samples would indicate excellent thermal transport 
properties due to the material hardness and purity. However,
typical device designs involve necessary nano\-structuring, for example, 
the construction of material heterostructures that define quantum wells 
or traps a transport channel below an oxide and above a doped substrate. 
The associated formation of interfaces is documented to effectively 
suppress the thermal transport both parallel
\cite{deyu,yao,tighe,lee,chen00,per1,ren,rennert,baladin,wire,wiretheory,X,richardson,zou,volz2,mingowire,luxiang,cher,X2,X3,barrat} and  
perpendicular~\cite{Schwab,chenperp,per2,per3,broidoperp,tamura,chen2,volz1,bies,chen0} to 
the materials boundaries due to the phonon Knudsen flow~\cite{chen00,per1,wire,Fuchs,hojgaard} 
and the phonon reflection~\cite{lee,ren,per2,per3,bies} effects, respectively.  
Both types of interface mechanisms lie outside the traditional Fourier 
analysis\cite{fourier} of heat conduction but produce a 
suppression of the effective thermal conductivity within a region extending 
to about one phonon mean-free path $\ell_{\rm mfp}$ from the interface.  
The interface impact of scattering can even cause crossovers in the
nanostructure conductivity such that a material with high bulk thermal
conductivity is not necessarily the best thermal conductor upon
nanostructuring.\cite{mingowire}  
While thermal-conductivity suppressions can work to improve the efficiency 
of both thermoelectric and thermoionic 
cooling,\cite{mahan2,Woods,hicksprb,mahan,sofo,reinecke,linchung,thermoelec} 
room-temperature estimates,\cite{elenipreprint} $\ell_{\rm mfp}(T=\hbox{300K})
\geq \hbox{100 nm}$, also testifies to some adverse effects even for
present devices.  Worse, the quest to nanosize modern electronics can only 
exacerbate the thermal dissipation problems\cite{cahill,cl_tien} since the increase 
in packing density simultaneously will produce a further degradation in the 
effective thermal-transport properties.

\begin{figure}[h]
\begin{center}
\hspace*{-1cm}
\includegraphics[width=8.5cm]{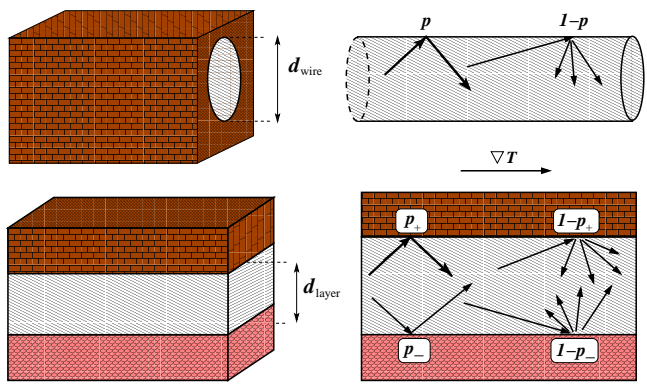}
\caption{\label{layer} 
Typical semiconductor nanostructures that are subject to a 
Knudsen-flow effect (Refs.~\protect\onlinecite{Fuchs} and 
\protect\onlinecite{hojgaard}) with a suppression of the thermal 
conductivity.  Phonon transport can be effectively confined to device channels 
(light gray) by the low conductivity of the processed surrounding.  
Moreover this in-channel thermal transport typically suffers a 
significant suppression by interface scattering.  The upper panel 
illustrates two possible realizations of a wire which could be 
embedded in an oxide or free-standing.  Phonons experience specular 
(diffusive) scattering with the probability $p$ ($1-p$) at the
wire boundaries.  The lower panel shows a corresponding realization 
of a multilayered nanostructure in which, for example, a transport
channel is located between two (in general) different materials 
(for example, a top oxide and a heavily doped substrate).  Phonons 
scatter specularly from the upper (lower) interface with probability 
$p_{+(-)}$ and diffusively with probability $1-p_{+(-)}$ to produce a 
Knudsen flow.}
\end{center}
\end{figure}

In this paper we illustrate the general nature of this Knudsen effect on the 
effective lattice thermal conductivity of semiconductor devices by documenting 
the effects on two simple sample structures: thin semiconducting 
wire\cite{deyu,wire,wiretheory,X,zou,volz2,mingowire,riccardo} and multilayered 
systems,\cite{yao,tighe,lee,chen1,chen00,per1,ren,rennert,baladin,chen0,elenipreprint}
Fig.~\ref{layer}.  
Semiconductor wires and layers represent examples of semiconductor 
nanostructures that are interesting in themselves. They can also be viewed 
as typical building blocks in general semiconducting devices that typically 
have transport channels sandwiched between an oxide and substrate.

Figure~\ref{layer} shows schematics of both types of nanostructures; 
the light regions in the schematics represent the 
transport active systems of high thermal conductivity while 
the dark regions illustrate the processed surrounding which 
may be oxides or heavily doped materials.  In both systems the 
oxide or the dopants will inhibit or limit the thermal transport in the 
processed regions and move the phonon transport to the transport 
active wire or layer. However, the interface scattering 
suppresses the thermal transport even in the (pure) transport 
channel where the scattering causes a Knudsen flow.\cite{Fuchs,hojgaard}

We obtain analytical results for the thermal-transport suppression given 
by special functions in both the wire and (multi)layer cases. 
The analytical results hold exactly when the phonon mean-free path
$\ell_{\rm mfp}$ can be approximated as constant\cite{generalize} 
and simplify the thermal-transport calculations for wires and 
multilayers to simple frequency integrals in the more general case 
when $\ell_{\rm mfp}$ depends on mode '$i$' and momentum '$q$'. We thus 
extend previous investigations\cite{yao,ren,per1,wire,chen00,chen0,zou,rennert,richardson} 
by efficient evaluations of the wire and layer thermal conductivities 
$\kappa_{\rm wire}$, $\kappa_{\rm layer}$, and by
(analytical) results for the layer Knudsen flow that arises
when the nature of interface scattering differs at the two 
materials boundaries in a multilayer structure. 
These results also permit an efficient evaluation of the 
multilayer thermal transport. 
We find that the Knudsen flow effect~\cite{hojgaard} is important in nanosized 
wires and multilayered structures formed from typical semiconductor materials 
due to the very large phonon mean-free paths.
We test the accuracy of the approximations used to calculate the phonon
Knudsen flow by comparing our results to experimental measurements for a
nanosized Si wire.\cite{deyu} 
We further investigate the phonon Knudsen flow in typical  
multilayer structures, including a simplified model of a nanosize transistor 
transport channel,\cite{elenipreprint} and demonstrate a breakdown of the 
traditional Fourier analysis\cite{fourier} for some semiconductor 
nanostructures at and below room temperature.  Finally we suggest 
an indirect but in-situ probe\cite{elenipreprint} for the oxidation 
progress in SiC (and Si) materials.

\begin{figure}[H]
\begin{center}
\includegraphics[width=17cm]{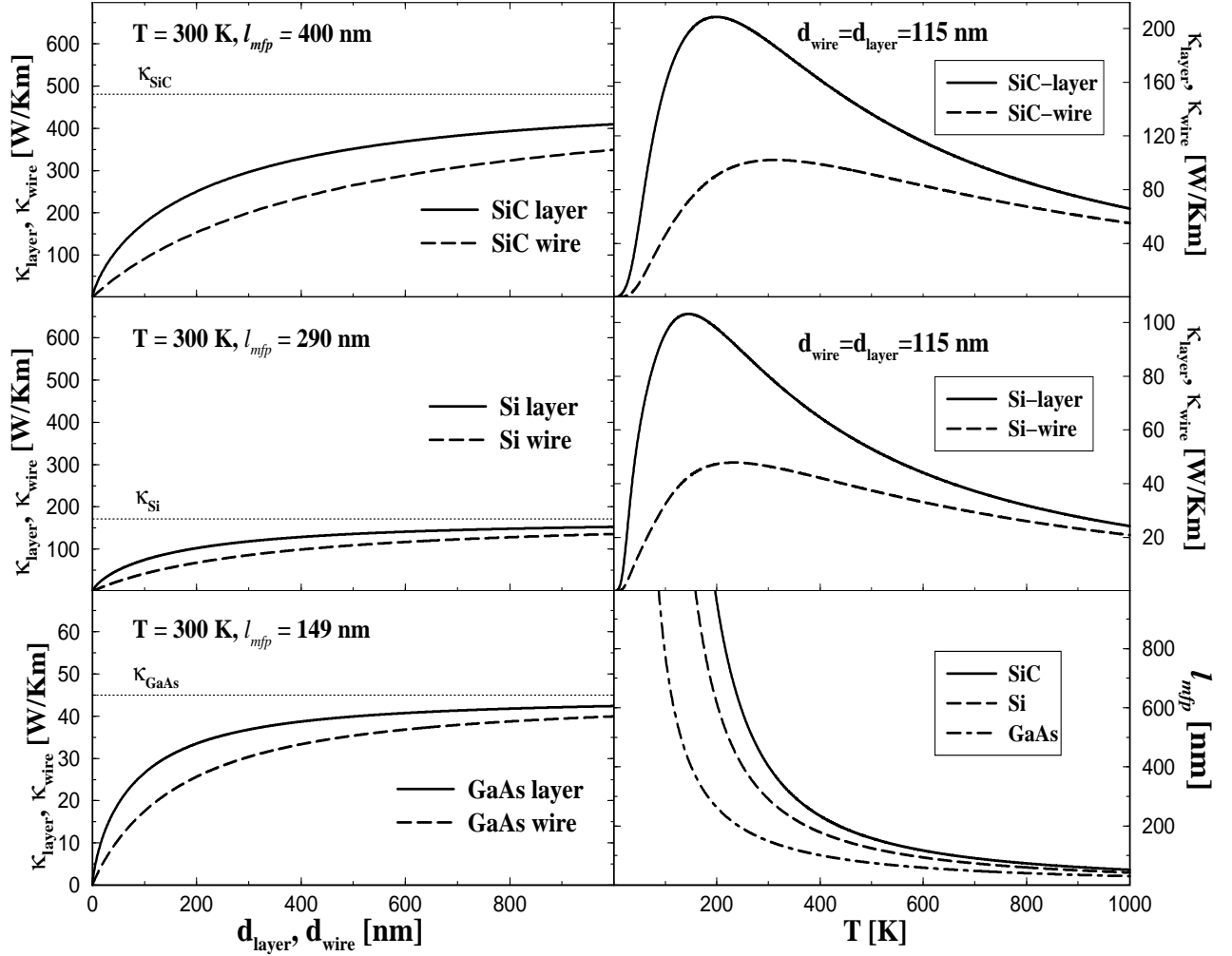}
\caption{\label{conductivity} 
Summary and origin of predicted suppression in the thermal 
conductivity of nanostrucured semiconductor structures.
{\it The left panels} show the thermal conductivity of layer 
(solid curves) and wire (dashed curves) structures calculated 
as a function of the layer thickness and wire diameter at 
room temperatures for three typical semiconductor materials.
The dotted lines identifies the experimentally observed value 
of the bulk-material thermal conductivities from which we also 
extract the listed values of the phonon mean free paths 
$\ell_{\rm mfp}$. A corresponding graph for Ge gives almost 
similar values to that of GaAs. {\it The right top} and 
{\it right middle} panels compare the predicted temperature 
variation of the layer and wire thermal conductivity in SiC 
and Si for a layer thicknesses $d_{\rm layer}$/wire diameter 
$d_{\rm wire}$ for which there now exist experimental realizations 
as Si nanowires~(Ref. \protect\onlinecite{deyu}). 
{\it Finally, the lower right} 
panel shows the calculated temperature variation in the almost 
micron-scale phonon mean free paths in bulk SiC, Si, and GaAs. 
The predicted phonon mean free path exceeds the typical separation 
between material interfaces in nanostructures and explains the device 
importance of the phonon Knudsen flow effect.}
\end{center}
\end{figure}

Figure~\ref{conductivity} summarizes the predicted 
suppression in the lattice thermal conductivity of general semiconductor 
wire and layer nanostructures.  We stress that there is a difference in 
thermal-conductivity range used to present the results for 
SiC, Si, and GaAs materials but that the overall physical behavior 
and consequences for semiconductor nanostructures remain unchanged 
in all materials.  The {\it left} panels show $\kappa_{\rm wire}$ 
(dashed curves) and $\kappa_{\rm layer}$ (solid curves) for varying 
thickness $d_{\rm wire,\,layer}$ for SiC, Si, and GaAs at $T=300$ K.
The dotted lines identify the experimentally observed value of 
the bulk-material thermal conductivities from which we also extract 
the listed values of the phonon mean-free paths $\ell_{\rm mfp}$.  
The {\it top right } and {\it middle right} panels document the
temperature variation in the Knudsen-flow thermal conductivity 
in SiC and Si layers (solid curves) and wires (dashed curves).
Finally, the {\it lower right} panel shows the calculated temperature 
variation in the phonon mean-free path $\ell_{\rm mfp}(T)$ 
for SiC, Si, and GaAs.  The phonon Knudsen-flow effect on the thermal 
transport arises because this $\ell_{\rm mfp}(T)$ approaches the microscale
at lower temperatures and may even at room temperature exceed
the typical separation between boundaries in modern and future semiconductor 
devices.

This paper is organized as follows. In section II we estimate 
the bulk-phonon mean-free path in SiC, Si, and GaAs, and 
we discuss the general Knudsen-flow behavior in low-dimensional systems.  
In section III we analyze the phonon-boundary scattering in wires and layered
structures and we present complete special-function 
evaluations\cite{elenipreprint,elenithesis,elenisolo} 
for the lattice thermal conductivities $\kappa_{\rm wire}$ and 
$\kappa_{\rm layer}$. In section IV we discuss and compare the present 
approach with traditional thermal transport calculations of low-dimensional 
systems, and we suggest a possible in-situ probe of the SiC oxidation process.
Finally, section V contains our conclusions.

\section{Phonon transport Theory}
We investigate the thermal transport properties in 
semiconductors where the dominant heat carriers
are the phonons.  For every phonon momentum $q$ (and 
phonon mode $i$) we determine the distribution function 
$N_{q(i)}$ which relaxes to an equilibrium value,  
$N^0_{q(i)}= 1/[\exp(\hbar \omega_q^{(i)}/k_B
      T)-1]$, given 
exclusively by the temperature $T$ and the phonon energy 
$\hbar \omega_q^{(i)}$ of said mode.  We use a relaxation time 
$\tau_{q}^{(i)}$ to describe the effective scattering in the bulk 
material. Further, we ignore the optical modes and assume for the 
longitudinal and transverse modes\cite{LB2,Feldman} ($i=l,t$) a simple 
isotropic phonon dispersion given by $q=\omega 
(1+\alpha_i\omega^3)/c_i$ and parameters $\alpha_i$ fitted to 
the phonon energies at the zone boundaries $\Theta_i$.\cite{ModeFit} 
Table I lists the sound velocities, $c_i$, and effective Debye 
temperatures, $\Theta_i$, for SiC, Si, GaAs, and Ge along with
the fitted values for the dispersion parameter $\alpha_i$. 

We mostly focus on the case with a constant, effective phonon 
mean-free path $\ell_{\rm mfp}$ (that only depends on temperature and 
materials) but we also provide results and analysis for the more 
general case of a (mode- and) momentum-dependent mean-free path
$\ell_{\rm mfp}^{(i)}(q)= {v}_{q}^{(i)}t_{q}^{(i)}$.
We use our model of the phonon dynamics to estimate 
the group velocities ${v}_{q}^{(i)} = c_i/(1+4\alpha_i\omega^3)$.  
In the following 
derivation and discussion we generally suppress the explicit labeling by 
mode index but all calculations have been performed by 
handling the longitudinal and transverse phonon modes separately.

The distribution change $\delta N_{q} = N_{q} - N_q^0$ 
induced by a small thermal gradient, $\nabla T$, 
is calculated by solving the 
Boltzmann transport equation in the relaxation-time 
approximation,\cite{boltzmann}
\begin{equation}
({\bf v}_q \cdot \nabla) N_{q} = -\frac{N_{q} - N_q^0}{\tau_q},
\label{eq:BTE}
\end{equation}
to linear order in the thermal gradient $\nabla T$.  The
$q$ (mode) dependent value of the relaxation rate $1/\tau_q$ 
is effectively fitted from bulk-thermal transport
measurements\cite{LB2,expcond} as 
discussed below. In turn, the calculated distribution change 
$\delta N_{q}$ permits our evaluation of the thermal current
\begin{equation}
\mbox{\boldmath $j$} =
\sum_i \int\frac{d^3q}{{(2\pi)}^3}
\hbar\omega_q^{i}{\bf v}_q^i\delta N_{qi},
\label{eq:heatcurrent}
\end{equation}
and hence of the thermal conductivity
\begin{equation}
\kappa=-\mbox{\boldmath $j$}{(\nabla T)}^{-1}.
\label{eq:fourier}
\end{equation}
The interface scattering enters from the proper inclusion of the
boundary conditions in Eq.~(\ref{eq:BTE}).  

\subsection*{II A. Estimate of the phonon mean-free path in bulk}

To obtain an approximate evaluation of the phonon mean-free path
we first consider the thermal transport in a bulk material where 
the solution of the linearized Boltzmann transport equation (l-BTE) 
simplifies to
\begin{equation}
\delta N_{q(i)}^{\rm Bulk} = -\tau_q^{(i)}
\left({\bf v}_q^{(i)}\cdot 
\nabla T\right)
\left(\frac{\partial N_{q(i)}^0}{\partial T}\right).
\end{equation}
The lattice thermal conductivity, $\kappa_{\rm Bulk}$, is calculated
by inserting this solution in Eqs.~(\ref{eq:heatcurrent}) 
and~(\ref{eq:fourier}).  For a frequency and polarization-independent 
phonon mean-free path, $\tau_q^{(i)}{v}_q^{(i)}\equiv \ell_{\rm mfp}(T)$, 
the lattice thermal conductivity in a bulk material $\kappa_{\rm Bulk}
\propto \ell_{\rm mfp}$ is simply proportional to this effective mean-free 
path.\cite{per1,wire} 
Comparison to experimental values for the
temperature variation in the bulk thermal conductivities, 
$\kappa_{\rm Bulk}^{\rm exp}(T)$, therefore yields a simple phonon 
mean-free path estimate\cite{per1,wire}
\begin{eqnarray}
\ell_{\rm mfp}(T) =  \kappa_{\rm Bulk}^{\rm exp}(T) \, \, \,
\frac{6\pi^2\hbar^3}{k_B^4T^3}
{\left(\frac{2f_t(T)}{c_t^{2}}+\frac{f_l(T)}{c_l^{2}}\right)^{-1}}.
\label{eq:mfp}
\end{eqnarray}
The estimate depends on the modal dispersion as expressed 
through the occupation-weighted integrals
\begin{equation}
f_i(T)=\int_0^{\Theta_i/T}dx\frac{x^4e^x}{{(e^x-1)}^2}
{\left[1+\alpha_i
{\left(\frac{xk_BT}{\hbar}\right)}^3\right]}^2.
\label{eq:f}
\end{equation}

The {\it lower right} panel of figure~\ref{conductivity} shows the 
calculated variation of the phonon mean-free path $\ell_{\rm mfp}$ as 
a function of the temperature for the set of semiconducting 
materials studied.  For GaAs and Ge we use values of 
$\kappa_{\rm Bulk}^{\rm exp}$ taken from Ref.~\protect\onlinecite{LB2} 
while for SiC and Si we use values from Ref.~\protect\onlinecite{expcond}.
The corresponding curve for Ge coincides almost exactly with 
that for GaAs because of the similarity in sound velocities 
and in measured values for the bulk thermal conductivities.
We note that the phonon mean-free paths in the three materials are 
almost of micron-scale dimensions and that SiC possesses the largest 
value of $\ell_{\rm mfp}$ ($\ell_{\rm mfp}^{\rm SiC} \approx 
1.5\,\ell_{\rm mfp}^{\rm Si} \approx 3\,\ell_{\rm mfp}^{\rm Ge,GaAs}$).

Our approximative constant-$\ell_{\rm mfp}$ description provides 
simple and analytical evaluations of the Knudsen thermal-transport 
suppression in both the wire and multilayer cases. This description 
permits qualitative estimates of the general effect.\cite{generalize}  
The actual bulk-phonon mean-free path $\ell^{(i)}_{\rm mfp}(q)= 
\tau_q^{(i)} {v}_q^{(i)}$ will, of course, depend on the mode 
and on the phonon momentum $q$ and/or the frequency 
$\omega_{q}^{(i)}$ and calls for a more refined 
analysis.
We test the validity of the constant-$\ell_{\rm mfp}$ results
by also considering the Knudsen flow arising with a frequency-dependent 
(mode-independent) effective decay rate~\cite{wire} $\tau = A(T)/\omega^{2}$
and by comparing the results of both theory descriptions to recent 
experimental observations of the thermal transport in Si 
nanowires.\cite{deyu}  For the test model we determine the 
function $A(T)$ from measurements~\cite{expcond} of the temperature 
variation of the bulk (Si) thermal conductivity using
\begin{eqnarray}
A(T) =  \kappa_{\rm Bulk}^{\rm exp}(T) \, \, \,
\frac{6\pi^2\hbar}{k_B^2T}
{\left(\frac{2g_t(T)}{c_t}+\frac{g_l(T)}{c_l}\right)^{-1}},
\label{eq:A_T}
\end{eqnarray}
where the functions $g_i$ are defined as
\begin{equation}
g_i(T)=\int_0^{\Theta_i/T}dx\frac{x^4e^x}{{(e^x-1)}^2}
\frac{{\left[1+\alpha_i
{\left(xk_BT/\hbar\right)}^3\right]}^2}
{\left[1+4\alpha_i
{\left(xk_BT/\hbar\right)}^3\right]}.
\label{eq:g}
\end{equation}
In this more refined model description we determine below the Knudsen 
suppression analytically up to a single integration over the
phonon frequency.

\subsection*{II B. Phonon Knudsen flow}
We focus our study of the Knudsen flow effect on two types of nanosizes 
semiconductor structures that can be viewed as general building blocks 
in devices. The {\it upper panel of Fig.~1} illustrates 
two possible realizations of wire systems~\cite{deyu,Schwab,wire,riccardo} 
which could be embedded in an oxide or could exists as a free-standing object.
The phonons are assumed to undergo either specular or diffusive scattering with 
the probability $p$ and $1-p$ at the wire boundary. The {\it lower 
panel of Fig.~1} shows a corresponding realization of a 
multilayer structure in which the transport-active channel 
(of high material purity) is located between two similar 
or different materials representing, for example, an oxide
cab and a highly doped substrate. The surrounding materials 
(and/or vacuum) suppress the average transport not only because 
they have inferior conductivities themselves but also because
their material boundaries cause a Knudsen-flow suppression 
even in the transport channel.  Again, two types of scattering events 
are possible and phonons are assumed to scatter specularly from the upper 
(lower) interface with probability $p_{+(-)}$ and diffusively with 
probability $1-p_{+(-)}$, respectively.  The value of $p_+$ will 
generally differ from $p_-$ as, for example, an oxide interface
is rough ($p_+=0$) whereas the interface to an underlying (doped
or undoped) substrate may be very smooth ($p_-\rightarrow 1$).

The phonon-boundary scattering off interface/surface 
defects becomes dominant when the phonon mean-free path exceeds 
the dimensions of the material.\cite{chen00,per1,wire,hojgaard}  
To analyze the general situation we introduce $q_\bot$ ($q_\|$)  to 
denote the component of the phonon momentum $q$ that is 
perpendicular to the material boundary $\Omega$ (parallel to 
the thermal gradient, $\nabla T$). Specular phonon scattering has 
only indirect effects on the overall transport 
because the phonons will have the same momentum component 
$q_\|$ in the direction of the thermal gradient $\nabla T$ 
as before the scattering event. For such specular scattering we have 
\begin{equation} 
\delta N_q(\Omega,\,q_\bot)=\delta N_q(\Omega,\,-q_\bot).
\label{eq:spec}
\end{equation} 
In contrast, diffusive interface scattering causes a significant 
transport suppression as the phonon emerges with a random direction 
given by the local thermal distribution. The implicit thermalization
ensures 
\begin{equation} 
\delta N_q(\Omega,\,q_\bot >0)=0,
\label{eq:diff}
\end{equation} 
and this boundary condition affects the distribution also 
away from the interface. 

To calculate the resulting impact on the thermal transport for our 
low-dimensional structures we solve the l-BTE 
\begin{eqnarray}
\left\{\frac{1}{q}\left(q_{\bot}\cdot \nabla \right)
+\frac{1}{\tau_q {v}_q}\right\} \delta N_q
=-\frac{1}{q}\,\frac{\partial N_q^0}{\partial T}\,
\left(q_\|\cdot \nabla T\right) 
\label{eq:BTEgeneral}
\end{eqnarray}
subject to the boundary conditions Eqs.~(\ref{eq:spec}) and (\ref{eq:diff}).
We express the general solution relative to the corresponding bulk case
\begin{equation} 
\delta N_q
=\delta N_{q}^{\rm Bulk}\left[1-h_{q}(r_\bot,q_\bot/q,p)\right]
\label{eq:gFunc}
\end{equation} 
using $r_\bot$ to denote the coordinate direction away from
the local surface. Furthermore, the form of the  
l-BTE~(\ref{eq:BTEgeneral}) permits us to express the relative
suppression in terms of an amplitude  $\phi_{q}(q_\bot/q,p)>0$ 
and an exponential decay
\begin{equation} 
h_{q}(r_\bot,q_\bot/q,p) =\phi_{q}(q_\bot/q,p)
e^{-r_\bot/\{\tau_q {v}_q (q_\bot/q)\}}.
\label{eq:hhFunc}
\end{equation} 
The amplitudes $\phi_{q}(q_\bot/q,p)$ are fixed by 
analysis of the boundary conditions.\cite{wire,Fuchs,elenithesis,elenisolo} 

\begin{figure}[h]
\begin{center}
\includegraphics[width=8.5cm]{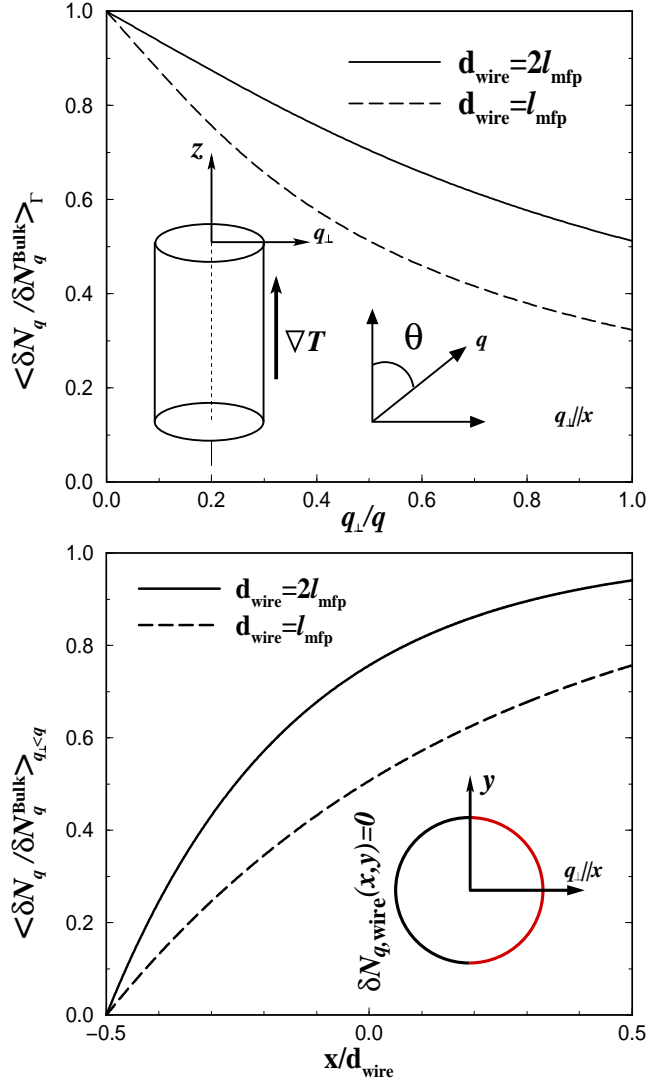}
\caption{\label{deviation} 
Schematics and details of the Knudsen-flow effect as expressed in the
momentum and spatial variation in the thermal-gradient induced
changes in the phonon distribution. The figure shows 
(for a given magnitude $q$ of the phonon momentum {\boldmath{$q$}}) 
the ratio of the distribution change 
$\delta N_q=\delta N_q^{\rm wire}$ (arising in a wire with purely diffusive 
interface scattering) to the corresponding bulk-distribution
changes $\delta N_q^{\rm Bulk}$ evaluated for different ratios of 
$d_{\rm wire} / \ell_{\rm mfp}(q)$. 
{\it The top panel} shows this distribution ratio 
(averaged over the wire cross section) as a function of the angel $\theta$ 
between the phonon momentum and the wire axis/thermal-gradient 
direction/transport direction $z$.
{\it The bottom panel} shows the corresponding variation of the 
distribution ratio (averaged over $\theta$) 
with the coordinate $x$ (at $y=0$) along the wire diameter which 
is assumed to coincide with the direction of the perpendicular momentum
component $q_\bot$.}
\end{center}
\end{figure}

Figure~\ref{deviation} illustrates the general nature of the Knudsen 
flow in the phonon distribution, $\delta N_q/\delta N_q^{\rm Bulk}$,
as functions of the momentum angle $q_{\bot}/q$ (top panel) and of the
spatial coordinate $x$ across the diameter (bottom panel). The Knudsen-flow
distribution $\delta N_q$ is specified by numerical 
integration~\cite{wire,rennert,zou} and/or analytical 
evaluation.\cite{per1,elenipreprint,elenithesis,elenisolo}  
The figure details the variation in $\delta N_{q}/\delta N_{q}^{\rm Bulk}$ and 
compares results obtained assuming purely diffusive scattering 
($p=0$) for two wires with diameters $d_{\rm wire}=
\ell_{\rm mfp}^{(i)}(q)$ and $2\ell^{(i)}_{\rm mfp}(q)$.
The calculated suppression applies for all phonon momenta 
(and modes) for wires when, as we typically do, we approximate 
the phonon mean-free path as constant. More generally, the figure 
illustrates the universal effect in the Knudsen flow with a suppression 
of distribution changes extending far away from the boundaries; similar 
momentum and spatial variation curves result both under more general 
assumptions for the mean-free path and in the case of multilayered 
semiconductor nanostructures.

The resulting effect on the thermal conductivity $\kappa$ can be summarized
by averaging the relative suppression $h_{q(i)}$ over the cross section 
(perpendicular to $\nabla T$). For every phonon $q$ and $i$ we define
$\mathcal{H}_{qi}(q_\bot/q,p) = \int_\Gamma d\Gamma \,h_{qi}(r_\bot,q_\bot/q,p)/\Gamma$
and express the nanostructure conductivity change 
\begin{eqnarray} 
\delta \kappa &=& \kappa - \kappa_{\rm Bulk}  
={\left(\nabla T\right)}^{-1} 
\sum_i\int \frac{d^3{q}}{{(2\pi)}^3}
\hbar \omega_{q}^i {v_q}^i
\,\frac{q_\|}{q}\,
\delta N_{qi}^{\rm Bulk}
\mathcal{H}_{qi}(q_\bot/q,p).
\label{eq:NOratio}
\end{eqnarray} 
It is convenient to introduce an effective inverse Knudsen number 
$\eta_i({q}) = d_{\rm wire(layer)}/(\tau_{q}^i{v_q}^i)$ 
and describe the thermal-conductivity suppression\cite{wire}
\begin{equation}
 -\frac{\delta \kappa}{\kappa_{\rm Bulk}}(p) = 
\frac{\sum_i \int_0^{\Theta_i/T}dx
\,{x^4e^x}\,{{(e^x-1)}^{-2}}\,
{\left[1+\alpha_i
{\left(xk_BT/\hbar\right)}^3\right]}^2
F_{qi}(p)|_{\eta_i(q(x))}}
{\sum_i \int_0^{\Theta_i/T}dx
\,{x^4e^x}\,{{(e^x-1)}^{-2}}\,
{\left[1+\alpha_i
{\left(xk_BT/\hbar\right)}^3\right]}^2}
\label{eq:NOratio2}
\end{equation}
as a weighted frequency integral over the effective suppression factor
\begin{equation}
F_{qi}(p)|_{\eta_i(q)}\equiv
\left(\frac{4\pi}{3}\right)^{-1}
\int_{|\vec{q}|=q} d^2 q \,\left(\frac{q_\|}{q}\right)^2 
\, \mathcal{H}_{qi}(q_\bot/q,p).
\label{eq:Fdefgen}
\end{equation} 
We emphasize that significant simplification follows from an assumption of a
mode and  frequency independent $\ell_{\rm mfp}={\rm constant}$.
In that case we obtain the simple result~\cite{chen00,per1,wire,zou}
\begin{equation} 
 -\frac{\delta \kappa}{\kappa_{\rm Bulk}}(\eta=\hbox{\rm \scriptsize constant},p) =
F_{qi}(p)|_{\eta_{i}(q)=\eta} 
\label{eq:F_H}
\end{equation} 
for the Knudsen effect on the nanostructure thermal transport.
This follows because the value of $\eta$ is then independent of $q$ and of
the mode.
For the case of wires and multilayered structures it is 
possible to obtain analytical results for the solution~(\ref{eq:F_H}).

\section{Interface-limited nanostructures thermal transport}

The study of the thermal transport in wires and layered structures
requires a careful analysis~\cite{Fuchs} of the phonon scattering 
at the interfaces of the specific systems. From this analysis we 
can complete the evaluation of the Knudsen transport
suppression~(\ref{eq:F_H}) for a free-standing wire, for a 
individual layer with possibly different nature of scattering
at the top and bottom interfaces, and for a multilayered structure.
Details of the derivation is published elsewhere~\cite{elenithesis,elenisolo} 
and here we just summarize the approach and the analytical results 
obtained.\cite{elenipreprint} 

\subsection*{III A. Thermal conductivity in wires}
For a free standing wire (Figure~\ref{layer}) we
assume a thermal gradient along the wire-axis '$z$'. For simplicity
we let the $x$-direction coincide with the perpendicular phonon
momentum, {\it i.e.,} $q_\bot = q_x$, and we assume the 
same specular scattering ratio $p$ along the whole 
boundary. Under these conditions we solve the l-BTE
(Eq.~(\ref{eq:BTEgeneral})) for the change $\delta N^{\rm wire}$ 
subject to the boundary conditions (\ref{eq:spec}) and 
(\ref{eq:diff}) with $\Omega_+ = \sqrt{(d/2)^2-y^2}$ for $x>0$
and $\Omega_- = -\sqrt{(d/2)^2-y^2}$ for $x<0$. We analyze the
boundary effect following Ref.~\protect\onlinecite{Fuchs}, 
adapting~\cite{wire,elenithesis} the description originally developed 
for electron transport in thin films.

Using the definition~\eqref{eq:gFunc} we formally determine the wire
suppression ratio
\begin{equation} 
h_{q}^{\rm wire}(x,\,y,q_x/q,p) = (1-p)
\frac{\exp(-\sqrt{(d/2)^2-y^2}/\{\tau_q v_q(q_x/q)\})}
{1-p\exp({-2\sqrt{(d/2)^2-y^2}/\{\tau_q v_q(q_x/q)\}})}
e^{-x/\{\tau_q v_q(q_x/q)\}}
\label{eq:hWire}
\end{equation}
and evaluate the average amplitude suppression~\cite{wire,elenithesis} 
\begin{eqnarray}
\mathcal{H}_q^{\rm wire}(q_\bot,/q,p) &=& \frac{4}{\pi}(1-p)^2\sum_{j=1}^\infty 
jp^{j-1} \int_0^1dt\;\sqrt{1-t^2}\;e^{-jt\eta(q)/(q_\bot/q)}.
\label{eq:wire1}
\end{eqnarray}
A polar coordinate system, where $q_x=q\sin\theta$, is then used  
to express the weighted Knudsen suppression factor \eqref{eq:Fdefgen} 
at mode {\it i},
\begin{eqnarray}
\left.F^{\rm wire}_{qi}(p)\right|_{\eta_i(q)}
&=&
{\left(\frac{4\pi}{3}\right)}^{-1}\int_0^{2\pi}d\varphi
\int_0^{\pi}d\theta\, \sin\theta \cos^2\theta
\,\mathcal{H}^{\rm wire}_{qi}(q_x/q=\sin\theta;\,p)
\nonumber\\[1mm]
&=&(1-p)^2\sum_{j=1}^\infty jp^{j-1}M(j\eta),
\label{eq:wire2}
\end{eqnarray}
in terms of a Meijer's $G$ special function~\cite{meijerg} and a polynomium:
\begin{equation}
M(\zeta) = \frac{6}{\zeta}
\left\{\frac{\pi^{3/2}}{4}G^{2 \,0}_{2 \,4}\left(
\begin{array}{c|llll}
\underline{\zeta^2} & 1,\,3 \\
4 & \frac{1}{2},\,\frac{3}{2},\,-\frac{1}{2},\,1
\end{array}\right)
-\left(\frac{\zeta^2}{6}-\frac{\zeta^4}{45}\right)\right\}. 
\label{eq:specfunc}
\end{equation}
For purely diffusive scattering ($p\rightarrow 0$) only the first term in the 
sum of Eq.~(\ref{eq:wire2}) contributes to the lattice thermal 
conductivity and therefore the ratio $\kappa_{\rm wire}/\kappa_{\rm Bulk}$ 
simplifies to  
\begin{equation} 
\frac{\kappa_{\rm wire}}{\kappa_{\rm Bulk}}(\eta,\,p\rightarrow 0)
=1-\frac{6}{\eta}\left\{\frac{\pi^{3/2}}{4}G^{2 \,0}_{2 \,4}\left(
\begin{array}{c|llll}
\underline{\eta^2} & 1,\,3 \\
4 & \frac{1}{2},\,\frac{3}{2},\,-\frac{1}{2},\,1
\end{array}\right)
-\frac{\eta^2}{6}+\frac{\eta^4}{45}\right\}.
\label{eq:constant_mfp}
\end{equation}
Similarly, if instead we assume pure specular scattering along the whole 
wire-boundary then 
$\left.F^{\rm wire}(\,p\rightarrow 1)\right|_\eta \rightarrow 0$ 
so that the bulk transport description is correctly recovered.

Our analytical results \eqref{eq:wire2}--\eqref{eq:specfunc} and/or \eqref{eq:constant_mfp} 
are in exact agreement with the numerical results\cite{wiretypo} presented in
Ref.~\protect\onlinecite{wire}. Our analytical l-BTE result differs
quantitatively but not qualitatively from the nonselfconsistent
$\kappa_{\rm wire}$ determination~\cite{richardson,richrefA} that 
can be extracted from analytical calculations of how boundary scattering 
shortens the average phonon mean-free path.\cite{richardson}  
We stress that our results  \eqref{eq:wire2}--\eqref{eq:specfunc} 
and/or \eqref{eq:constant_mfp} completes the previously published 
investigation~\cite{wire} by a special function evaluation for a
significant speed up in computation efficiency. We also stress that 
care must be taken in the numerical evaluation of our special 
function~(\ref{eq:specfunc}), or the integral in Eq.~\eqref{eq:wire2} 
for large arguments ($\zeta > 10$) when using, for example, the Mathematica 
program.

\subsection*{III B. Thermal conductivity in an individual layer}
We consider a layer of thickness $d$ with its boundaries 
at $z = \pm d/2$ and a thermal gradient lying in the $xy$-plane.
In this geometry we have $q_\bot = q_z$.  In the general case, 
$p_+\neq p_-$, we must analyze the phonon scattering at both 
interfaces in order to evaluate the induced change, 
$\delta N_{q}^{\rm layer}$.
Generalizing previous results~\cite{chen00,per1,wire,Fuchs,hojgaard} 
and introducing $\eta=d_{\rm layer}/\ell_{\rm mfp}$ and
$\xi = \eta /(2|q_z|/q)$, we obtain for general $p_{+/-}$ the
result~\cite{elenithesis,elenisolo}
\begin{equation} 
\left(\begin{array}{c}
\phi_{q,q_z>0}^{\rm layer}(p_+,\,p_-) \\ \\
\phi_{q,q_z<0}^{\rm layer}(p_+,\,p_-)
\end{array}\right)
=\frac{1}{e^{2\xi} - p_-p_+e^{-2\xi}}
\left(\begin{array}{c}
(p_--1)e^{\xi}+p_-(p_+-1 )e^{-\xi}\\ \\
(p_+-1 )e^{\xi}+p_+(p_--1 )e^{-\xi}
\end{array}\right),
\label{eq:philayer}
\end{equation} 
for the amplitudes of the suppression caused by scattering at the upper 
and lower interfaces.

For our single-layer and multilayer calculations we will generally assume that 
we have a constant $\ell_{\rm mfp}$ and can hence express the Knudsen
transport suppression 
\begin{eqnarray} 
\frac{\kappa_{\rm layer}}{\kappa_{\rm Bulk}}(\eta,\,p) &=& 
1-F^{\rm layer}(p_+,p_-)|_{\eta =\hbox{\rm \scriptsize constant}};
\label{eq:Layer_cond}
\end{eqnarray} 
generalizations to a mode and frequency-dependent phonon
mean-free path are as straightforward as in the wire case.
We calculate again the cross-section average of the distributions 
change, $\mathcal{H}_{q}^{\rm layer}(q_z/q;p_+,p_-)$, and determine 
the phonon suppression factor as an average over the direction of momentum
$q$ 
\begin{eqnarray}
\left.F^{\rm layer}(p_+,p_-)\right|_{\eta =\hbox{\rm \scriptsize constant}}
&=& 
\frac{3}{4}
\int_{0}^\pi  d\theta\sin^3\theta
\mathcal{H}_{q}^{\rm layer}\left(\frac{q_\bot}{q}=\cos\theta;\,p_+,\,p_-\right),
\label{eq:Flayer}
\end{eqnarray} 
noting that the sign of $\cos(\theta)=q_\bot/q=q_z/q$ uniquely determines
whether the emerging phonons have scattered off the top or bottom
material boundary.\cite{elenithesis} Introducing
finally
\begin{equation} 
\bar{p} = \frac{p_++p_-}{2}\quad {\rm and} \quad \hat{p} = \sqrt{p_+p_-}.
\label{eq:pLayer}
\end{equation} 
and performing the $\theta$-integration we obtain the analytical 
result\cite{elenipreprint,elenithesis,elenisolo} 
\begin{eqnarray}
\left.F^{\rm layer}(\bar{p},\hat{p})\right|_{\eta=\hbox{\rm \scriptsize constant}}&=& \frac{3}{8\eta}
\left\{(1-\bar{p})-4\sum_{j=1}^\infty Q(j,\,\eta,\,\bar{p},\,\hat{p})
\right\},
\label{eq:Flayer2}
\end{eqnarray}
given in terms of a modified exponential error function  
\begin{eqnarray} 
Q(j,\,\eta,\,\bar{p},\,\hat{p}) = 
(\hat{p}^2-2\bar{p}+1)\hat{p}^{j-1}D(j\eta)+(\bar{p}-\hat{p})(\hat{p}+1)^2
\hat{p}^{2(j-1)}D(2j\eta),
\label{eq:Qfunc}
\end{eqnarray}
where
\begin{eqnarray} 
D(a\eta) &=& 
\left(\frac{1}{4}-5\frac{a\eta}{12}-\frac{(a\eta)^2}{24}+\frac{(a\eta)^3}{24}\right)e^{-a\eta}
\nonumber\\[3mm]
&+&\left(\frac{(a\eta)^2}{2}-\frac{(a\eta)^4}{24}\right)\int_{\eta}^{\infty}dk\;\frac{e^{-ak}}{k}.
\label{eq:layer}
\end{eqnarray}
At $p_+ = p_-$ the present more general result~(\ref{eq:Flayer2})-(\ref{eq:layer}) 
correctly simplifies to
\begin{eqnarray} 
\frac{\kappa_{\rm layer}}{\kappa_{\rm Bulk}}(\eta,\,p) &=& 1-(1-p)\frac{3}{8\eta}
\left\{1-4(1-p)\sum_{j=1}^\infty p^{j-1}D(j\eta)
\right\},
\label{eq:Pequal}
\end{eqnarray}
as obtained in previous derivations.\cite{chen00,per1,wire,chen2}

\begin{figure}[h]
\begin{center}
\includegraphics[width=8.5cm]{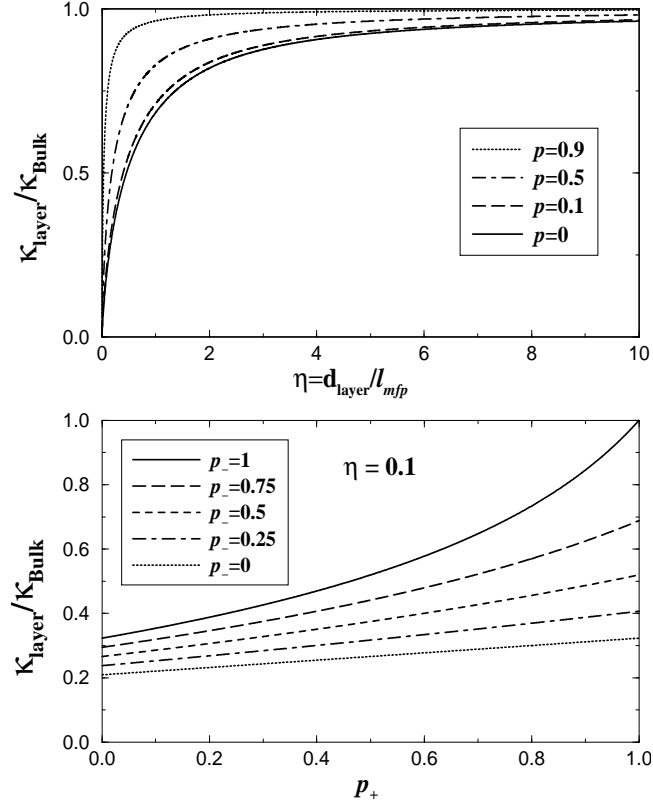}
\caption{\label{spec} Relative suppression of the thermal conductivity in 
a general individual semiconductor layer evaluated for a the constant mean free
path $\ell_{\rm mfp}$.  {\it The upper panel} shows 
the thermal conductivity ratios,  $\kappa_{\rm layer}/
\kappa_{\rm Bulk}$, as a function of $\eta$ for specular scattering 
probabilities $p_+ = p_- = p$.  The thermal conductivity naturally 
increases to recover the bulk-like behavior recovers when the 
probability for specular scattering increases.  {\it The lower 
panel} shows the ratio $\kappa_{\rm layer}/\kappa_{\rm Bulk}$ 
as a function of $p_+$ for different $p_-$ evaluated at a given ratio $\eta =
d_{\rm layer}/\ell_{\rm mfp} = 0.1$.
The panel documents that the layer thermal 
conductivity remains significantly suppressed from the bulk 
transport value even when the phonons scatter purely 
specularly ($p_\pm=1$) from one of the interfaces.} 
\end{center}
\end{figure}

Figure~\ref{spec} shows the reduction of the lattice thermal conductivity 
in pure semiconductor layers. 
The {\it top} panel documents the ratio $\kappa_{\rm layer}/\kappa_{\rm Bulk}$ 
as a function of $\eta\equiv d_{\rm layer}/\ell_{\rm mfp}$ 
for $p_+ = p_- = p$.
As in the wire case, the lattice thermal conductivity 
naturally becomes higher to recover the bulk behavior 
when the probability for specular scattering increases. 
The {\it lower} panel shows $\kappa_{\rm layer}/\kappa_{\rm Bulk}$ 
for a fixed layer-thickness $d_{\rm layer} = 0.1\ell_{\rm mfp}$ and for 
varying specular scattering probabilities, $p_+$ and $p_-$.
The figure documents that the thermal conductivity remains suppressed 
below the bulk transport value even when the phonons are 
purely specularly scattered from one of the two interfaces.

\subsection*{III C. Thermal transport in multilayered structures}

To estimate the thermal transport in heterostructures   
we make the approximation to ignore all phonon transmission between
the layers. There exists exact results~\cite{rennert} but our focus is on
providing simple estimates and extracting consequences of the size-limited
thermal transport.
We consider structures with layers $j=1,\,2\ldots$ of individual
thicknesses $d_j$ and with probabilities $p_{+(-)}^{j}$ for 
specular (rather than diffusive) scattering at the top (bottom) 
layer boundary. Our approximation then consists in approximating 
the effective (linear-response) conductivity as
\begin{equation}
\kappa_{\rm eff}=\frac{\sum_j \kappa_j(d_j,p_+^j,p_-^j) \, d_j}{\sum_j d_j},
\label{eq:multilayer}
\end{equation}
where $\kappa_j(d_j,p_+^j,p_-^j)$ denotes the thermal conductivity 
of layer $j$.  
We stress that the individual-layer thermal conductivity typically is
subject to a significant Knudsen-flow suppression and must be calculated for 
the layer-specific nature of the boundary scattering ($p_{+/-}^j$)
and for the specific individual-layer thickness $d_j$, as also 
indicated by the argument to the individual-layer conductivities $\kappa_j$. 

\section{Results and discussion}
The Knudsen flow effect has direct consequences for the semiconductor
nanostructure thermal transport. Below we test the accuracy of our
approximations and detail the size effects on the temperature variation
of $\kappa_{\rm wire}$ and $\kappa_{\rm layer}$.

\subsection*{IV A. Thermal transport in Si and SiC nanostructures}
\begin{figure}[h]
\begin{center}
\includegraphics[width=8.5cm]{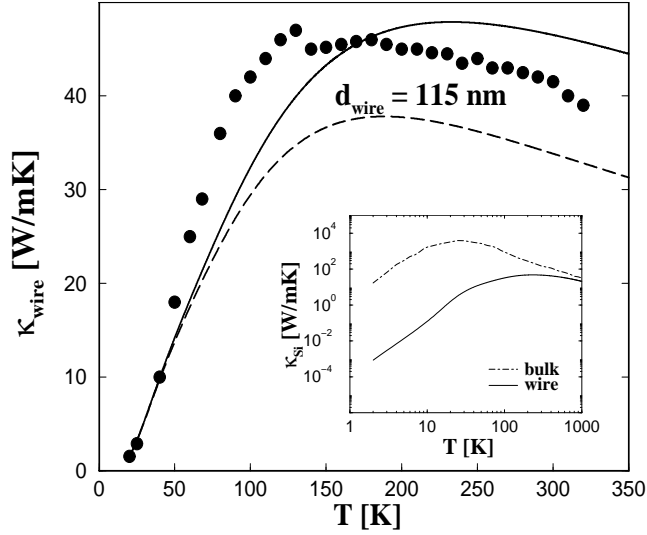}
\caption{\label{fig:Si_T} Test of accuracy of our approximative phonon Knudsen flow
  description. The figure shows the temperature variation of the lattice 
  thermal conductivity in a 115 nm thick silicon wire. The solid circles 
  represent experimental values from Ref.~\protect\onlinecite{deyu}. 
  The dashed (solid) curve shows our theory estimate for pure
  diffusive scattering, $p=0$, and calculated assuming a 
  frequency-dependent relaxation time~\cite{wire} $\tau_q = A(T)/\omega_q^{2}$
  (a frequency-/polarization--independent bulk-phonon mean 
  free path $\ell_{\rm mfp}(T)$ specified by bulk-measurements.) 
  The inset compares the temperature-variation on logarithmic scales 
  of bulk-Si thermal transport (dashed-dotted curve) and the evaluated
  Si-wire conductivity (solid curve) calculated for a frequency-independent 
  $\ell_{\rm mfp}(T)$ .}
\end{center}
\end{figure}

Figure~\ref{fig:Si_T} tests the accuracy of our approximative determination 
of the Knudsen flow effects by comparing the measured\cite{deyu} and  
calculated temperature variation of the lattice  thermal conductivity 
in a 115 nm thick silicon wire. 
The solid circles represent experimental data from 
Ref.~\protect\onlinecite{deyu} while the solid (dashed) curve shows
our theoretical results  \eqref{eq:wire2}--\eqref{eq:specfunc}, \eqref{eq:constant_mfp}
for the Knudsen flow calculated for pure diffusive scattering ($p=0$) and assuming a 
frequency and polarization-independent bulk-phonon mean-free path $\ell_{\rm mfp}$ 
(a frequency-dependent relaxation time~\cite{wire} $\tau_q =
A(T)/\omega_q^{2}$ using (\ref{eq:A_T})$-$(\ref{eq:g}) to calculate the
function $A(T)$). 
The insert compares the predicted Si wire thermal conductivity with
measurements~\cite{expcond} of bulk Si on logarithmic scales, emphasizing the
significant effects of nanostructuring.

The figure reveals some discrepancies which we assign to the simple
approximation we use for the phonon dispersion.\cite{wiretheory} 
The figure also shows that the more elaborate model with a 
frequency dependent $\tau$ and $\ell_{\rm mfp}$  yields the more 
accurate description.  However, since these two models give 
similar magnitude and qualitative behavior of 
$\kappa_{\rm wire}$ (with good agreement to 
the measured values) we find that the analytical 
result~\eqref{eq:constant_mfp} serves as a simple, 
analytical estimate of the wire conductivity
$\kappa_{\rm wire}$.

\begin{figure}[h]
\begin{center}
\includegraphics[width=8.5cm]{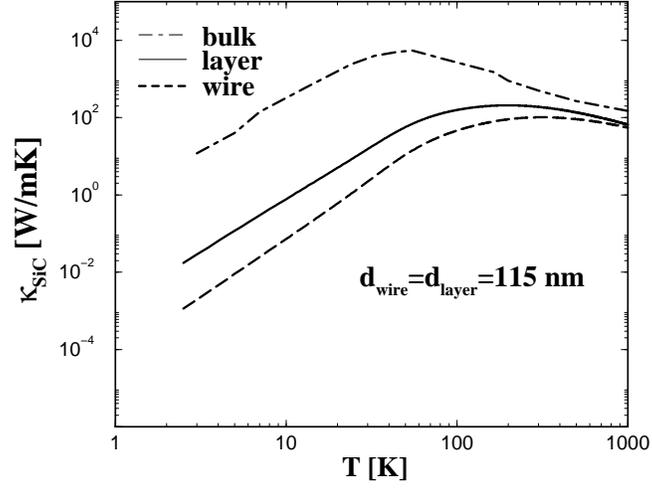}
\caption{\label{fig:SiC_T} Thermal conductivity suppression in 115 nm thick
  SiC wires (dashed curve) and layers (solid curve) compared to the bulk-SiC
  thermal conductivity measurements (dashed-dotted curve).\cite{expcond}  
  The peak values for the wire and layer transport case are about two orders
  of magnitude smaller than the bulk-transport peak and are shifted to much
  higher temperatures.}
\end{center}
\end{figure}

Figure~\ref{fig:SiC_T} compares the corresponding thermal-conductivity
variation with temperature of bulk SiC (dashed-dotted curve) and of a
115 nm thick SiC wire (dashed curve) and layer (solid curve) structure. 
For both nanostructures we assume pure diffusive
scattering ($p=0$) at all interfaces and a frequency and 
mode-independent bulk-phonon mean-free path described by Eq.~(\ref{eq:mfp}). 

Together the insert of Fig.~\ref{fig:Si_T} and Fig.~\ref{fig:SiC_T} 
stress the two features which are important when studying the thermal transport in
nanostructure (Si and SiC) systems:\cite{deyu,wiretheory} 
(i) the maximum bulk-transport value
is about two orders of magnitude higher than the corresponding peak values
for the nanostructures, {\it i.e.,} 
$\kappa_{\rm Bulk}^{\rm max}(T) \approx 100\kappa_{\rm layer,\,wire}^{\rm max}(T)$, 
and (ii) the maximum values of the lattice thermal
conductivity in the wires and layers are shifted towards higher temperatures
from the peak of the bulk conductivities.

\subsection*{IV B. Thermal transport in doped materials}

\begin{figure}[h]
\begin{center}
\includegraphics[width=17cm]{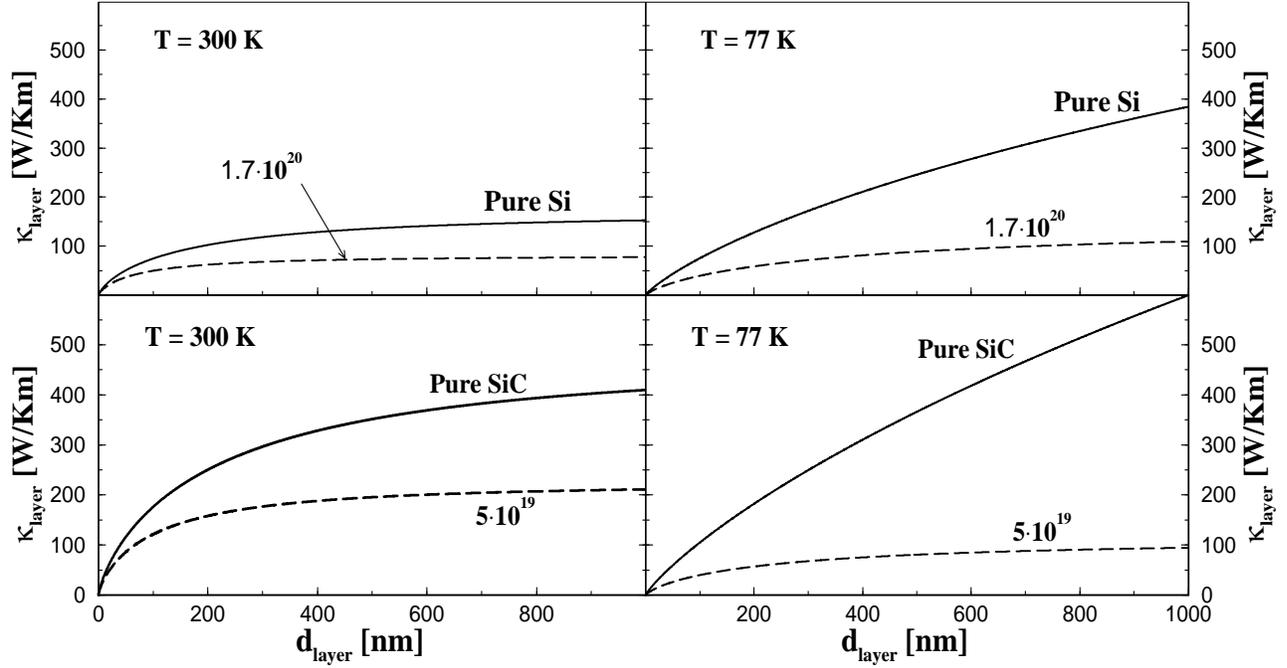}
\caption{\label{Layer} Reduction of the lattice thermal 
conductivity in doped and undoped layers of Si (top panels) and SiC (bottom panels).
In all four panels the lattice thermal conductivity, $\kappa_{\rm layer}$, is
plotted as a function of the layer-thickness, 
$d_{\rm layer}$, for $p_+=p_-=0$ at $T=300$ K (left panels)
and $T=77$ K (right panels).
The {\it top panels} show  $\kappa_{\rm layer}$ in pure (solid curves)
and in p-doped (dashed curves) Si layers.  Similarly, the {\it bottom panels} 
compare the lattice thermal conductivity in pure SiC layers (solid curves) to 
that of n-doped layers (dashed curves).}
\end{center}
\end{figure}

Figure~\ref{Layer} documents that the Knudsen suppression in the 
lattice thermal conductivity also affects the case of doped of 
Si and SiC nanostructures.  The figure compares the conductivity 
of a Si and SiC layer at two different temperatures, 77 K and 300 K 
for pure materials (solid curves) and in the case of a very high 
dopant concentration $\sim 10^{20}$ cm$^{-3}$  (dashed curves).
We assume here that the phonons are scattered purely diffusively
at both interfaces ($p_+=p_-=0$) and use the analytical result
\eqref{eq:Flayer2}--\eqref{eq:layer} for an assumed constant  
$\ell_{\rm mfp}(T)$.  
We use the measurements of Ref.~\protect\onlinecite{expcond} with the Si
layer p--doped to be $1.7\cdot 10^{20}$ cm$^{-3}$ and the SiC layer n--doped 
$5\cdot 10^{19}$ cm$^{-3}$. The phonon mean-free path in the doped layers is 
thus estimated at room temperature to be 
49 nm and 138 nm for SiC and Si, respectively and at $T=77$ K to be 
420 nm and 830 nm, respectively.

\subsection*{IV C. Breakdown of Fourier transport analysis}
\begin{figure}[h]
\begin{center}
\includegraphics[width=8.5cm]{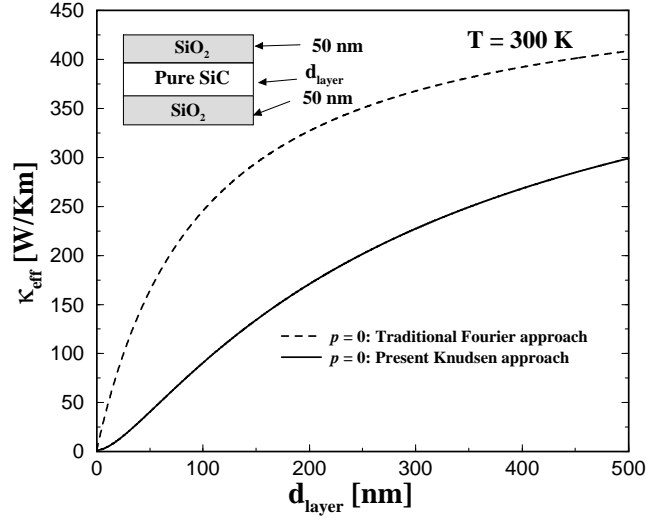}
\caption{\label{Layer2} Comparison of traditional thermal-transport 
calculations\protect\cite{fourier} (dashed curve) and present 
Knudsen-flow results (solid curve), Eqs.~(\ref{eq:multilayer}) and
(\ref{eq:Layer_cond})--(\ref{eq:layer}). 
The panel shows the effective lattice thermal conductivity 
as a function of the layer thickness in a SiC layer that is 
located between two 50 nm thick SiO$_2$ layers.
The traditional thermal-transport calculations overestimates 
the thermal conductivity significantly in low-dimensional 
semiconductor structures at room temperatures and below.
The figure therefore suggests a breakdown of the Fourier 
approach based on bulk conductivities as in typically used 
in finite-element calculations.}
\end{center}
\end{figure}

Figure~\ref{Layer2} documents that traditional thermal transport 
calculations fail to accurately describe the phonon transport in 
low-dimensional systems at and below room temperature. 
This is illustrated for a SiO$_2$/SiC/SiO$_2$ heterostructure for 
which we have calculated the effective multilayer thermal 
conductivity~\eqref{eq:multilayer} for varying thickness of 
the SiC layer, $d_{\rm SiC}$. At the SiC/SiO$_2$ interface 
we assume pure diffusive phonon scattering ($p=0$).
The solid curve represents the effective lattice thermal conductivity 
obtained by the present Knudsen approach (described in
Sec. III B) in which the thermal-conductivity suppression in each
individual layer is considered. In contrast, the dashed curve was obtained 
by replacing the suppressed value of the layer conductivity in 
Eq.~(\ref{eq:multilayer}) by the corresponding bulk transport conductivity,
as is typically done in many finite element calculations.  This Fourier 
analysis yields an effective lattice thermal conductivity which is too 
large.  Traditional Fourier analysis~\cite{fourier} of the lattice thermal 
transport is inadequate for some nanostructured semiconductor systems in 
the low-temperature regime and the Knudsen-flow suppression affects even 
the room-temperature behavior.

\subsection*{IV D. SiC oxidation}
\begin{figure}[h]
\begin{center}
\includegraphics[width=8.5cm]{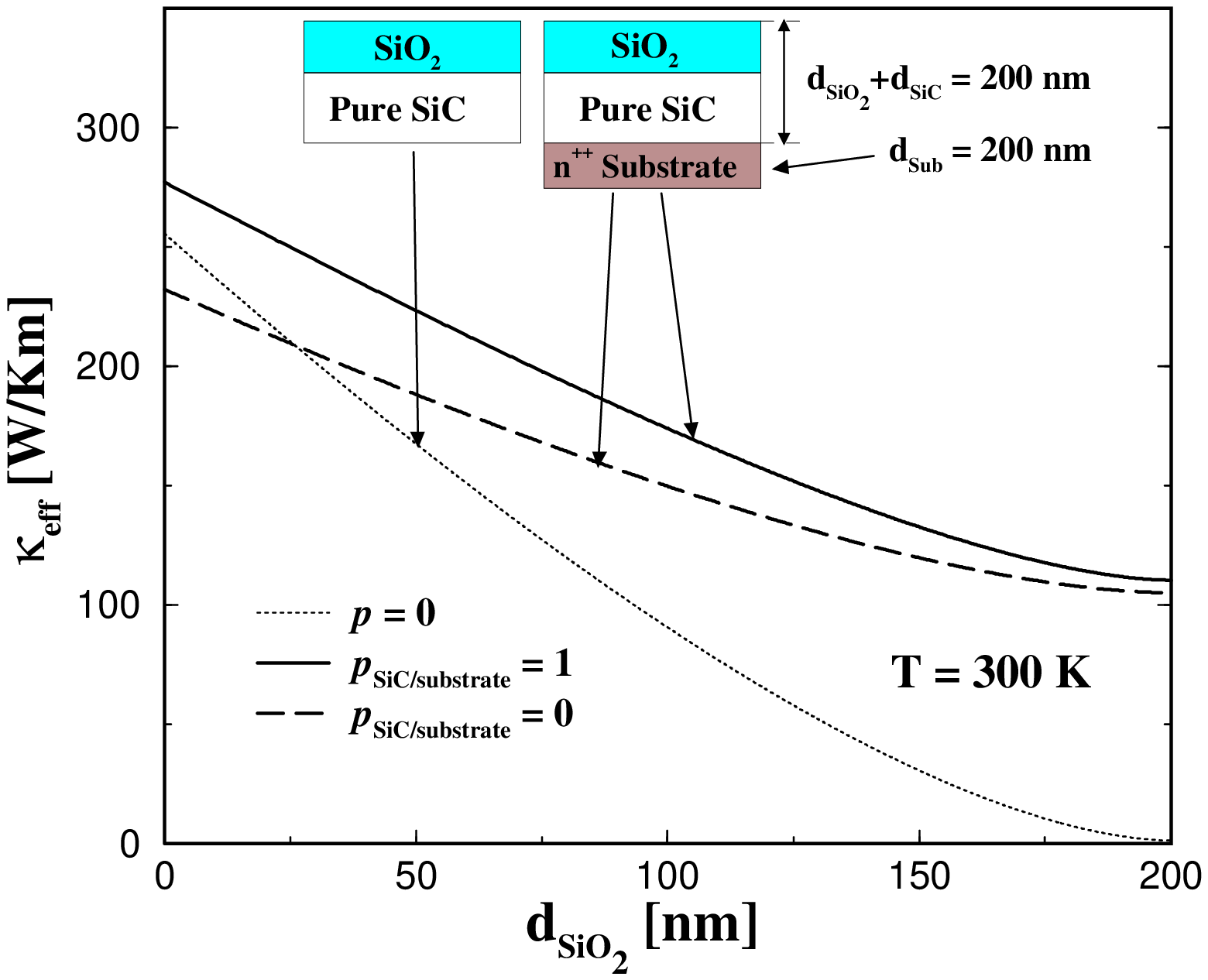}
\caption{\label{Oxide} Possible probe of the SiC oxidation 
process as measured by the variation in the effective lattice 
thermal conductivity with the oxide-thickness, $d_{\rm SiO_2}$.
The dotted curve correspond to a single SiC layer on which a thin 
uniform SiO$_2$ film grows.
The solid (dashed) curve shows results for a doped-SiC/SiC/SiO$_2$
heterostructure evaluated under the assumption that $p=1$ ($p=0$) 
characterize the scattering at doped-SiC/SiC interface and $p=0$
characterize the SiO$_2$/SiC interface. We assume 
that the underlying SiC layer is doped at a concentration of 
$5\cdot 10^{19}$ cm$^{-3}$.  For simplicity we also assume that
the thickness $d_{\rm SiO_2}+d_{\rm SiC}=200$ nm is kept fixed 
during the whole oxidation process that we 
model\protect\cite{footnote}.} 
\end{center}
\end{figure}

Figure~\ref{Oxide} illustrates the use of the Knudsen--flow thermal
transport as a possible gauge of the SiC oxidation process. 
Metal-oxide-semiconductor transistor gates are usually formed
by oxidizing the Si to form a thin oxide film~\cite{tan} located on 
top of the pure transport channel. 
This has been the focus of many experimental and theoretical
investigations.\cite{oxide1,oxide2,oxide3,oxide4,riccardo2,elwira}
Many of the experimental probes require that one extracts the sample and
investigate it in, for example, a transmission electron microscope whereas
qualitative in-situ probes could be more useful for some applications. 
We note that the SiC/SiO$_2$ interface is rough but of relative uniform
thickness.\cite{riccardo2} The situation is similar to the case of
adsorption on metal films where the (electrical) conductive is proven to
be sensitive~\cite{poland,film1,film2,film3,film4,poland2} to the film
thickness and surface nature.
We therefore argue~\cite{elenipreprint} that measurements of the thermal
Knudsen flow could serve as such an in-situ probe of the oxidation 
progress.\cite{oxide1,oxide2,oxide3,oxide4,riccardo2,elwira} 

The figure shows the variation of the effective lattice thermal conductivity, 
$\kappa_{\rm eff}$, with the oxide-thickness, $d_{\rm SiO_2}$, calculated by
Eq.~\eqref{eq:multilayer}. We use the value~\cite{ieee} 1.4 W m$^{-1}$ K$^{-1}$ 
for the bulk thermal conductivity of SiO$_2$ to calculate the 
effective lattice thermal conductivity in two different heterostructures. 
In addition, we assume a uniform oxide-thickness during the whole
oxidation process.
The dotted curve corresponds to a single SiC layer on which a thin uniform SiO$_2$
film grows. For simplicity we assume that $p=0$ at the SiO$_2$/SiC interface
and that the thickness $d_{\rm SiO_2}+d_{\rm SiC}=200$ nm
is kept fixed during the whole oxidation process that we model.\cite{footnote} 
The solid (dashed) curve correspond to results for a doped-SiC/SiC/SiO$_2$
heterostructure evaluated under the assumption that $p=1$ ($p=0$) 
characterizes the scattering at the doped-SiC/SiC interface; we also assume that
the underlying SiC layer is doped at a concentration of $5\cdot 10^{19}$
cm$^{-3}$. Our results document that the effective lattice thermal
conductivity is sensitive to the oxide thickness and could be used as a 
qualitative in-situ probe of the SiC oxidation progress.\cite{elenipreprint}

\section{Conclusions}
We have theoretically investigated the in-plane thermal transport
in nanostructured semiconductor wires, layers and multilayers,
in the framework of the l-BTE. The calculations include 
the important boundary scattering which specifies the phonon Knudsen 
flow and plays a dominant role in determining the thermal transport 
in semiconductor nanostructures. We obtain analytical results for the
nano\-structure thermal conductivity under the assumption of a frequency
independent (constant) mean-free path $\ell_{\rm mfp}$. Our results
further simplifies the thermal-transport calculations under more 
accurate assumptions for $\ell_{\rm mfp}$.  We find that the lattice 
thermal conductivity is significantly suppressed from the bulk value at 
room temperature and below in both thin-wire and thin-layer nanostructures.

We have tested our approximative description of the phonon Knudsen flow
effect by comparison to measurements of the thermal conductivity for 
thin Si wires and find that our analytical result for a constant 
$\ell_{\rm mfp}$ provides a fair accuracy in a qualitative description.
We further show that the traditional Fourier analysis~\cite{fourier} 
(typically used in finite-element calculations) predict an effective 
lattice thermal conductivity which is inaccurate for some nanostructured 
semiconductor systems especially in the low-temperature regime. 
Finally we have illustrated the use of the Knudsen-suppression thermal
transport as a possible gauge of the SiC oxidation process.

\section*{Acknowledgments}

We thank the National graduate school in materials science, the EU project 
ATOMCAD, and the Swedish Foundation for Strategic Research (SSF) though 
ATOMICS for support.

\newpage

\centerline{TABLES}

\begin{table}[h]
\begin{center}
\caption{\label{table0} 
Model parameters of the phonon description in SiC, Si, GaAs, and Ge.
The measured longitudinal and transverse sound velocities, $c_l$ and $c_t$, 
in SiC are taken from Ref.~\protect\onlinecite{Feldman} while 
for Si, GaAs, and Ge we use Ref.~\protect\onlinecite{LB2}.
The Debye temperatures, $\Theta_l$ and $\Theta_t$, 
are defined as $\Theta_i = \hbar \omega_i(q_{\rm max})/k_B$ 
and have in all four cases been calculated from the phonon dispersion curves in 
Ref.~\protect\onlinecite{LB2}.}
\begin{ruledtabular}
\begin{tabular}{lcccccc}
 Material& $c_l$ & $c_t$& $\Theta_l$& $\Theta_t$ & $\alpha_{l}\omega_l^3(q_{\rm max})$ &
$\alpha_{t}\omega_t^3(q_{\rm max})$ \\
\hline
SiC& 13300 & 7250 & 880 & 380 & 0.72 & 1.23 \\
\hline
Si& 8480 & 5870 & 570 & 160 & 0.43 & 1.91 \\
\hline
Ge& 4960 &3570 & 360 & 120 & 0.20 & 1.51 \\
\hline
GaAs& 4730 & 3340 & 270 & 100 & 0.55 & 1.95 \\
\end{tabular}
\end{ruledtabular}
\end{center}
\end{table}

\newpage

\centerline{FIGURES}

\begin{thebibliography}{99}

\bibitem{mahan2} G. Mahan, B. Sales, and J. Sharp, Phys. Today {\bf 50}, 
42 (1997).

\bibitem{cahill} D. G. Cahill, W. K. Ford, K. E. Goodson, G. Mahan, A. 
Majumdar, H. J. Maris, R. Merlin, and S. R. Phillpot, J. Appl. Phys. {\bf
  93}, 793 (2003).

\bibitem{deyu} D. Li, Y. Wu, P. Kim, L. Shi, P. Yang, and A. Majumdar, 
Appl. Phys. Lett. {\bf 83}, 2934 (2003).

\bibitem{yao} T. Yao, Appl. Phys. Lett. {\bf 51}, 1798 (1987).

\bibitem{tighe} T. S. Tighe, J. M. Worlock, and M. L. Roukes, 
  Appl. Phys. Lett. {\bf 70}, 2687 (1997).

\bibitem{lee} S.-M. Lee, D. G. Cahill, and R. Venkatasubramanian,
  Appl. Phys. Lett. {\bf 70}, 2957 (1997).  

\bibitem{chen1} X. Y. Yu, G. Chen, A. Verma, and J. S. Smith, 
Appl. Phys. Lett. {\bf 67}, 3554 (1995).

\bibitem{cahill2} D. G. Cahill, A. Bullen, and S.-M. Lee, 
High Temperatures--High Pressures {\bf 32}, 134 (2000).

\bibitem{Schwab} K. Schwab, E.~A.~Henriksen, J.~M.~Worlock,
and M.~L.~Roukes, Nature {\bf 404}, 974 (2000).

\bibitem{chen00} G. Chen, J. Heat Transfer Trans. {\bf 119},
  220 (1997).

\bibitem{per1} P. Hyldgaard and G. D. Mahan, 
{\it Thermal conductivity} (Technomic, Lancaster, PA 1996) {\bf 23}, 172 (1996).

\bibitem{ren} S. Y. Ren and J. Dow, Phys. Rev. B {\bf 25}, 3750 (1982).

\bibitem{rennert} P. Rennert and A. Brzezinski, Phys. Rev. B {\bf 52}, 1612
  (1999).

\bibitem{baladin} A. Baladin and K. Wang, Phys. Rev. B {\bf 58}, 1544
  (1998).

\bibitem{wire} S. G. Walkauskas, D. A. Broido, and K. Kempa,  
J. Appl. Phys. {\bf 85}, 2579 (1999).

\bibitem{wiretheory}  N. Mingo, L. Yang, D. Li, and
A. Majumdar, Nanoletters, {\bf 3}, 1713 (2003).

\bibitem{X} X. L\"{u}, W. Z. Shen, and J. H. Chu, J. Appl. Phys. {\bf 91},
  1542 (2002). 

\bibitem{richardson} R. A. Richardson and F.~Nori, Phys. 
  Rev. B {\bf 48}, 15209 (1993).

\bibitem{zou} J. Zou and A. Baladin, J. Appl. Phys. {\bf 89}, 2932 (2001). 

\bibitem{volz2} S. G. Volz and G. Chen, Appl. Phys. Lett. {\bf 75}, 2056
  (1999). 

\bibitem{mingowire} N. Mingo and D. A. Broido, Phys. Rev. Lett. {\bf 93},
  246106 (2004).

\bibitem{luxiang} L. Xiang, G. Ji-Hua, and C. Jun-Hao, Chinese Physics {\bf
    10}, 223 (2001).

\bibitem{cher} A. I. Chervanyov, Phys. Rev. B {\bf 66}, 214302 (2002).

\bibitem{X2} X. L\"u, J. H. Chu, W. Z. Shen, J. Appl. Phys. {\bf 93}, 1219
  (2003).  

\bibitem{X3} X. L\"u and J. Chu, Phys. Stat. Sol. (b) {\bf 226}, 285 (2001).     

\bibitem{barrat} P. Chantrenne and J-L. Barrat, J. Heat Transfer Trans. {\bf 126},
  577 (2004).

\bibitem{chenperp} G. Chen,  Phys. Rev. B {\bf 57}, 14958 (1998).

\bibitem{per2} P. Hyldgaard and G. D. Mahan, Phys. Rev. B {\bf 56}, 10754
  (1997).

\bibitem{per3} P. Hyldgaard, Phys. Rev. B {\bf 69}, 193305 (2004).

\bibitem{broidoperp} D.A. Broido and T. L. Reinecke, Phys. Rev. B {\bf 70},
  081310 (2004). 

\bibitem{tamura} S. Tamura, Y. Tanaka, H. J. Maris, Phys. Rev. B {\bf 60},
  2627 (1999).

\bibitem{chen2} G. Chen and M. Neagu, Appl. Phys. Lett. {\bf 71}, 2761
  (1997). 

\bibitem{volz1} S. Volz, J-B. Saulnier, G. Chen, and P. Beauchamp, 
High Temperatures--High Pressures {\bf 32}, 709 (2000).

\bibitem{bies} W. E. Bies, H. Ehrenreich, and E. Runge, J. Appl. Phys. 
{\bf 91}, 2033 (2002). 

\bibitem{chen0} G. Chen, T. Zeng, T. Borca-Tasciuc, and D. Song, Materials
  Science and Engineering {\bf A292}, 155 (2000).

\bibitem{cl_tien} C. L. Tien and G. Chen, J. Heat Transfear {\bf 116}, 799
  (1994). 

\bibitem{Fuchs} K. Fuchs, Proc. Cambridge Philos.~Soc.~{\bf 34}, 100
  (1938).

\bibitem{hojgaard} H. Smith and H. H\o jgaard Jensen, {\it Transport Phenomena}
  (Clarendon press, Oxford, 1989) pp. 246--257. 

\bibitem{fourier} E. Kreyszig, {\it Advanced Engeneering Mathematics} 
  8th Ed (Wiley, New York, 1999), ch 10.

\bibitem{Woods} G. D. Mahan and L. M. Woods, Phys. Rev. Lett. {\bf 80}, 4016
  (1998). 

\bibitem{hicksprb} L. D. Hicks and M. S. Dresselhaus, Phys. Rev. B {\bf 47}, 
  12727 (1993).

\bibitem{mahan} G. D. Mahan and H. B. Lyon, Jr., J. Appl. Phys. {\bf 76},
  1899 (1994).

\bibitem{sofo} J. O. Sofo and G. D. Mahan,  Appl. Phys. Lett. {\bf 65}, 2690
  (1994). 

\bibitem{reinecke} D. A. Broido and T. L. Reinecke, Phys. Rev. B {\bf 51},
  13797 (1995). 

\bibitem{linchung} P. J. Lin-Chung and T. L. Reinecke, Phys. Rev. B {\bf 51}, 
  13244 (1955).

\bibitem{thermoelec} D. A. Broido and T. L. Reinecke, Appl. Phys. Lett. 
{\bf 67}, 1170 (1995). 

\bibitem{elenipreprint} E.~Ziambaras and P. Hyldgaard, A brief report
  of selected multilayer transport results has been 
  submitted as a conference proceedings; {\tt cond-mat/0407122}. 

\bibitem{riccardo} R. Rurali and N. Lorente, Phys. Rev. Lett. {\bf 94},
  026805 (2005).

\bibitem{generalize} In fact, our analytical results for the 
wire and layer/multilayer Knudsen-flow suppression apply for 
transport of general quasiparticles with an isotropic dispersion 
assuming that the quasiparticles can be ascribed a constant mean 
free path. Our analytical results should thus be directly applicable 
also for electronic transport in, for example, thin simple- and 
noble-metal films and wires.

\bibitem{elenithesis} E.~Ziambaras, Licentiate thesis, Chalmers University of
Technology, (2004).

\bibitem{elenisolo} E. Ziambaras, in preparation. 

\bibitem{LB2} 
{\em Landolt-B\"ornstein: 
Numerical Data and Functional Relationships in Science and Technology,}
edited by O. Madelung, New Series, Group III, Vol. 41A1a 
(Springer, Berlin, 2001).

\bibitem{Feldman} D. W. Feldman, J. H. Parker, W. Choyke, and L. Patrick,
Phys. Rev.  {\bf 173}, 787 (1968). 

\bibitem{ModeFit} The parameters $\alpha_i$ are obtained from 
$\alpha_i = (q_{\rm max}c_i/\omega(q_{\rm max}) -1)/\omega^3(q_{\rm max})$ to fit the energies 
at the zone boundaries $\Theta_i = \hbar\omega(q_{\rm max})/k_B$ which are
estimated from the phonon dispersion curves in Ref.~\protect\onlinecite{LB2}. 

\bibitem{boltzmann} J. M. Ziman, {\it Electrons and Phonons} 
(Oxford University Press, Oxford, 1960), p. 264.

\bibitem{expcond} G. A. Slack, J. Appl. Phys. {\bf 35}, 3460 (1964).

\bibitem{meijerg}  A. P. Prudnikov, Yu. A. Brychkov, O. I. Marichev, 
{\it Integrals and Series}, Vol. 3: More special functions, 
(Gordon and Breach, New York, 1990).

\bibitem{wiretypo} Our result \eqref{eq:specfunc} disagrees, however, with
  the formal result, Eqs.~(7)--(9), of Ref.~\protect\onlinecite{wire} and 
we suspect a simple typo in their Eq.~(7) that yields negative results
for the thermal conductivity in very thin wires. 

\bibitem{richrefA} M.~I.~Flik and C.~L.~Tien, J.~Heat Transfer {\bf 112}, 
872 (1990).

\bibitem{tan} J. Tan, M. K. Das, J. A. Cooper, Jr. and M. R. Melloch,
  Appl. Phys. Lett. {\bf 70}, 2280 (1997).

\bibitem{oxide1} M. DiVentra and S. Pantelides, Phys. Rev. Lett. {\bf 83}, 
  1624 (1999).

\bibitem{oxide2} A. Gali, D. Heringer, P. De\'ak, Z. Hajnal, Th. Frauenheim,
  R. P. Devaty, and W. J. Choyke, Phys. Rev. B {\bf 66}, 125208 (2002).

\bibitem{oxide3} M. DiVentra and S. Pantelides, Phys. Rev. Lett. {\bf 86}, 
  5946 (2001).

\bibitem{oxide4} P. De\'ak, A. Gali, Z. Hajnal, Th. Frauenheim, N. T. Son,
  E. Janz\'en, and W. J. Choyke, Mat.~Sci.~Forum {\bf 433-436}, 535 (2003).

\bibitem{riccardo2} R. Rurali, E. Wachowicz, P. Ordejon, P. Godignon,
  J. Rebollo, and P. Hyldgaard, Mat.~Sci.~Forum {\bf 457-460}, 1293 (2004). 

\bibitem{elwira} E. Wachowicz, R. Rurali, P. Ordejon, and P. Hyldgaard,
  Comp. Mat. Sci. {\bf 33}, 13 (2005).

\bibitem{poland} R. Du\'s, E. Nowicka, and R. Nowakowski, Langmuir {\bf 20},
  9138 (2004).

\bibitem{film1} G. Palasantzas, Y.-P. Zhao, G.-C. Wang, T.-M. Lu, J. Barnas,
  and J. Th. M. De Hosson, Phys. Rev. B {\bf 61}, 11109 (2000).


\bibitem{film2} E. Z. Luo, S. Heun, M. Kennedy, J. Wollschl\"ager, and
  M. Henzler, Phys. Rev. B {\bf 49}, 4858 (1994).

\bibitem{film3} S. Kumar and G. C. Vradis, J. Heat Transfer Trans. {\bf 116},
  28 (1994).

\bibitem{film4} H. Grabhorn, A. Otto, D. Schumacher, and B.N.J. Persson,
  Surf. Sci. {\bf 264}, 327 (1992).

\bibitem{poland2} R. Du\'s, private communication.

\bibitem{ieee} M. G. Burzo, P. L. Komarov, and P. E. Raad,
IEEE Trans.~Comp.~Pack.~Tech.~{\bf 26}, 80 (2003).

\bibitem{footnote} In typical experiments the final oxide thickness 
is about two times the starting silicon thickness, Ref.~\protect\onlinecite{tan}, 
but that complication has no effect on the present conclusions due to the
vanishing contribution of the SiO$_2$ to the overall thermal conduction.


\end{thebibliography}
\end{document}